\documentclass{emulateapj}

\shorttitle{Evidence of a Metal Rich Galactic Bar}
\shortauthors{Soto et al.}

\begin{document}

\title{Evidence of a Metal Rich Galactic Bar from the Vertex Deviation of the
Velocity Ellipsoid }

\author{M. Soto }
\affil{Leiden Observatory, Leiden University, P.O. Box 9513, 2300RA, Leiden, The Netherlands}
\email{soto@strw.leidenuniv.nl}

\author{R.M. Rich } 
\affil{Department of Physics and Astronomy, UCLA, Los Angeles, CA, 90095-1547}
\email{rmr@astro.ucla.edu}

\author{K. Kuijken }
\affil{Leiden Observatory, Leiden University, P.O. Box 9513, 2300RA, Leiden, The Netherlands}
\email{kuijken@strw.leidenuniv.nl}

\begin{abstract}
We combine radial velocities, proper motions, and low resolution abundances
for a sample of $315$ K and M giants in the Baade's Window $(l,b)=(0.9^\circ,-4^\circ)$
Galactic bulge field.    The velocity ellipsoid of stars with $\rm [Fe/H]>-0.5$ dex 
shows a vertex deviation in the plot of
radial versus transverse velocity, consistent with that expected
from a population with orbits supporting a bar.   We demonstrate that
the significance of this vertex deviation using non-parametric
rank correlation statistic is $>99$\%.  The velocity
ellipsoid for the metal poor ($\rm [FeH]<-0.5$) part of the population shows no
vertex deviation and is consistent with an isotropic, oblate rotating
population.   We find no evidence for kinematic subgroups, but there
is a mild tendency for the vertical velocity dispersion $\sigma_b$ to 
decrease with increasing metallicity.

\end{abstract}


\section{Introduction}

The galactic bulge is the nearest example of a bulge/spheroidal
population.  A factor of 100 closer than M31, the proximity of the
bulge not only permits detailed measurement of stellar abundances and
radial velocities, but also of stellar proper motions.  The first such
study undertaken (at the suggestion RMR) was that of \citet{sp92}.
The proper motions were measured from plates obtained at the 200-inch
prime focus in 1950 and 1983; these data produced measurable proper
motions on the order of $0.5 \rm \ mas yr^{-1}$.

The \citet{sp92} study was the source of an input catalog for a
subsequent study of bulge radial velocity and low resolution
abundances \citep{terndrup95,sadler96}.  An early study of the
abundances and kinematics of these stars, using published abundances
from \citet{rich88} and radial velocities from \citet{rich90} was
undertaken by \citet{zhao94}.  That study suffered from having a
relatively small sample size, and the lack of a definitive bulge iron
abundance scale; nonetheless the most metal rich subset of stars
exhibited a vertex deviation of the velocity ellipsoid in the plot of
radial versus transverse velocity.  Modeling by \citet{zhao94} showed
that such a vertex deviation, or rotation of the major axis of the
velocity ellipsoid, is a feature expected from the long axis orbits
that would support a bar structure. Such a bar had been claimed by
\citet{BlitzSpergel} based on asymmetries in surface photometry of the
bulge region, and by \citet{binney} from the kinematics of molecular gas.

The possibility of a dominant stellar bar population was strengthened
by the spectacular results from the {\sl COBE} satellite
\citep{dwek94}.  All-sky maps of the Milky Way showed an asymmetry
that was consistent with a bar, with major axis in the first
quandrant.  Many subsequent studies confirm the presence of a bar
structure based on deprojecting the spatial distribution of the
stellar population \citep{bin97} and the Schwarzschild method has been
used to build a self consistent rapidly rotating bar \citep{zhao96,
  hafner00}.

Here we combine the proper motions of \citet{sp92}, the abundances of
\citet{terndrup95}, and the radial velocities of \citet{sadler96} to
explore the correlation of abundances and kinematics in the bulge
stellar population.  Surveys now in progress \citep{sumi04} have
achieved total numbers of order $5\times 10^6$ stars.  However, the
sample of 315 stars on which we report here is at present the largest
that has proper motions, radial velocities, and spectroscopically
estimated abundances.

In exploring correlations of abundances and kinematics, our aim is to
constrain models for the origin of the bulge and bar.  When the
composition of bulge stars \citep{mr94,fmr06b} is considered, it can
be argued that the bulge formed early and rapidly \citep{matt99,
  ferr03}.  \citet{kr02} isolated the old bulge population by proper
motion, and \citet{zoc03} demonstrated a comparison with a 13 Gyr old
main sequence turnoff population; these studies find little room for a
detectable trace young population in the bulge.  However, there are
complicating issues in the chemical evolution.  Mg is found to be elevated in
bulge stars, while O is only mildly so \citep{mr94,fmr06b,lecu06} and the
"explosive alphas" Ca, Si, and Ti are also less enhanced than Mg,
tracking the thick disk trends with [Fe/H].  Finally, the bar poses a
problem for early formation and enrichment, as the longevity of bars
is an issue.  If the metals in the bulge were built early and rapidly,
how and when did the bar appear and why has it survived?

Correlating abundances and kinematics gives another viewpoint on
chemical evolution.  Even if the abundance distribution appears to fit
the closed box model and the population is old, the correlation of
abundances and kinematics may reveal distinct kinematic subgroups.
Adding two proper motion vectors to the radial velocity increases the
power of this analysis.  There have been indications of
abundance/kinematics correlations in the bulge.  \citet{rich90} found
that the most metal rich stars have a smaller velocity dispersion.
\citet{sp92} found a similar trend in their small sample.
\citet{zhao94} found evidence for vertex deviation in the most metal
rich stars.  In a radial velocity study of a large number of bulge M
giants, \citet{scw90} found indications for a smaller velocity
dispersion for a metal rich subgroup of stars.  \citet{min96} obtained
spectroscopy in a bulge field at (l,b)=(8,7) and found differing
kinematics (more rotation support and a colder velocity dispersion)
for stars with $\rm [Fe/H]>-1$.  In this paper, we explore
abundance/kinematics correlations using both the radial velocity and
proper motion data, and abundances from low resolution spectroscopy.
We use the new iron abundance scale of \citet{fmr06a} to recalibrate
the \citet{sadler96} low resolution abundances (those being based on
the average of the Fe 5270 and 5335 features in the spectra).  While
the derived abundance for any individual star is uncertain to roughly
$+0.3$ dex, the calibration ensures that the kinematic transitions we
report here are tied to the best currently available iron abundance
scale.  The most metal rich stars in \citet{sadler96} have formal
metallicities in excess of 1 dex; we do not take those values
seriously, as they are transformed from outliers.
    
In this {\it Letter} we describe our kinematic analysis of the sample
of 315 red giants.

\section{Analysis}

\begin{figure*}
\includegraphics[width=0.32\hsize]{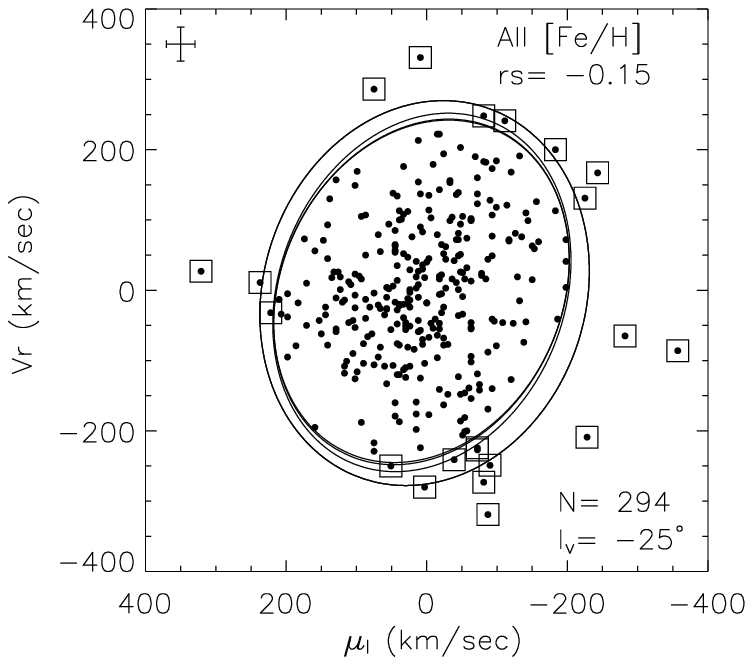}
\includegraphics[width=0.32\hsize]{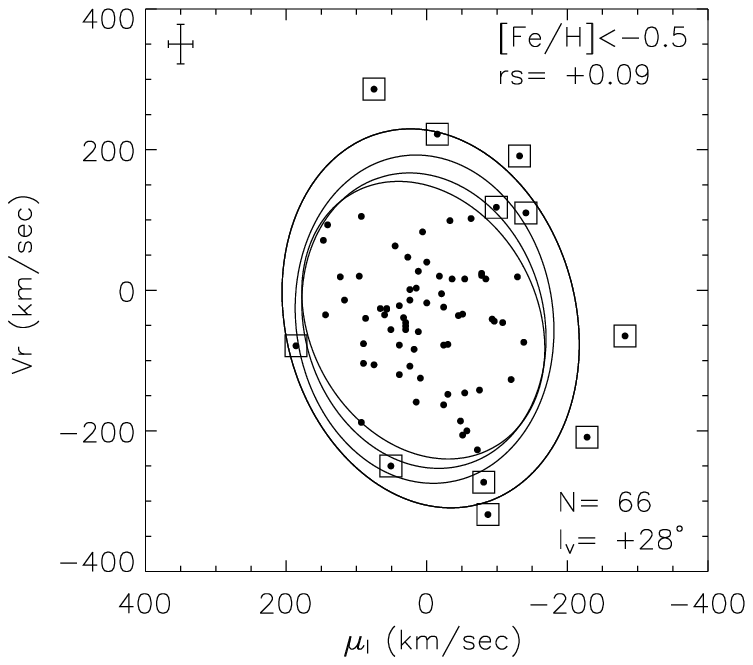}
\includegraphics[width=0.32\hsize]{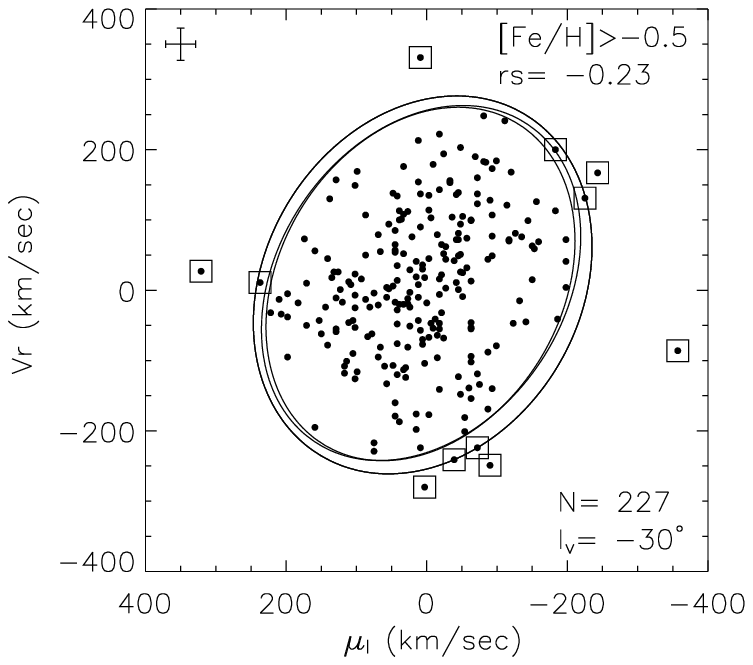}
\caption{Velocity ellipsoids for metal rich and metal poor populations in 
 Baade's Window. An iterative clipping algorithm to reject stars outside the 
 velocity ellipsoid at 2.5 $\sigma$ has been applied.
\emph{Left to right}, full sample, stars with $\rm [Fe/H]<-0.5$, and stars
with $\rm [Fe/H]>-0.5$. In each panel the vertex deviation $l_v$, the
Spearman rank correlation coefficient $r_S$, and the number of stars $N$,
is listed.
 \label{fig:veldisp}}
\end{figure*}

In an axisymmetric potential, one of the axes of the velocity
dispersion tensor of an equilibrium population must lie in the
meridional plane. Any deviation from this situation indicates a
non-axisymmetry, or a time variation. We will use the space velocities
of our stars to measure the orientation of the axis of the velocity
dispersion tensor in the radial-longitudinal velocity plane (known as
the {\em vertex deviation $l_v$}) and determine whether it is
consistent with zero.  The angle $l_v$ is defined as
\begin{equation}
\tan(2l_v) = \frac{2\sigma_{lr}^2}{|\sigma_r^2 - \sigma_l^2|}
\end{equation}
where $\sigma_{lr}$, $\sigma_r$ and $\sigma_l$
are the covariance and standard deviation of the measured velocity
components along the $r$ (line of sight) and $l$ (galactic longitude)
directions.  We measure the dispersions with an iterative clipping
algorithm, by repeatedly rejecting stars that lie outside the ellipse
with principal axes of 2.5 times the corresponding measured dispersion.
Table~\ref{tab:veldisp} and Fig.~\ref{fig:veldisp} summarize the
velocity ellipsoid parameters found, both for the full sample and for
metal poor and metal rich subsets. For each subsample we have
calculated the Spearman rank correlation coefficient $r_s$ and its
significance.

\begin{table}
\begin{center}
\caption{Velocity ellipsoid parameters.\label{tab:veldisp}. 
}
\begin{tabular}{ l  c  c  c  c  c  c  c  c }
\tableline\tableline
[Fe/H]            & $\sigma_r$   &   $\sigma_l$  & $\sigma_b$ &  $l_v$ &
$r_S$  &  N (N$_{\rm rej}$) &  Prob $r_S$ \\
                 &  \multicolumn{3}{c}{km/sec}    &  deg  & &   & \% \\
\tableline
All        & 109 & 94 & 86 & -25 & -0.15  & 315(21) &  0.85 \\ 
$<$-0.5    & 108 & 85 & 94 &  28 & +0.09  &  77(11) &  39 \\ 
$>$-0.5    & 107 & 96 & 83 & -30 & -0.23  & 238(11) &  0.02 \\

\tableline
\end{tabular}
\end{center}
\end{table}

\begin{figure*}
\includegraphics[width=0.32\hsize]{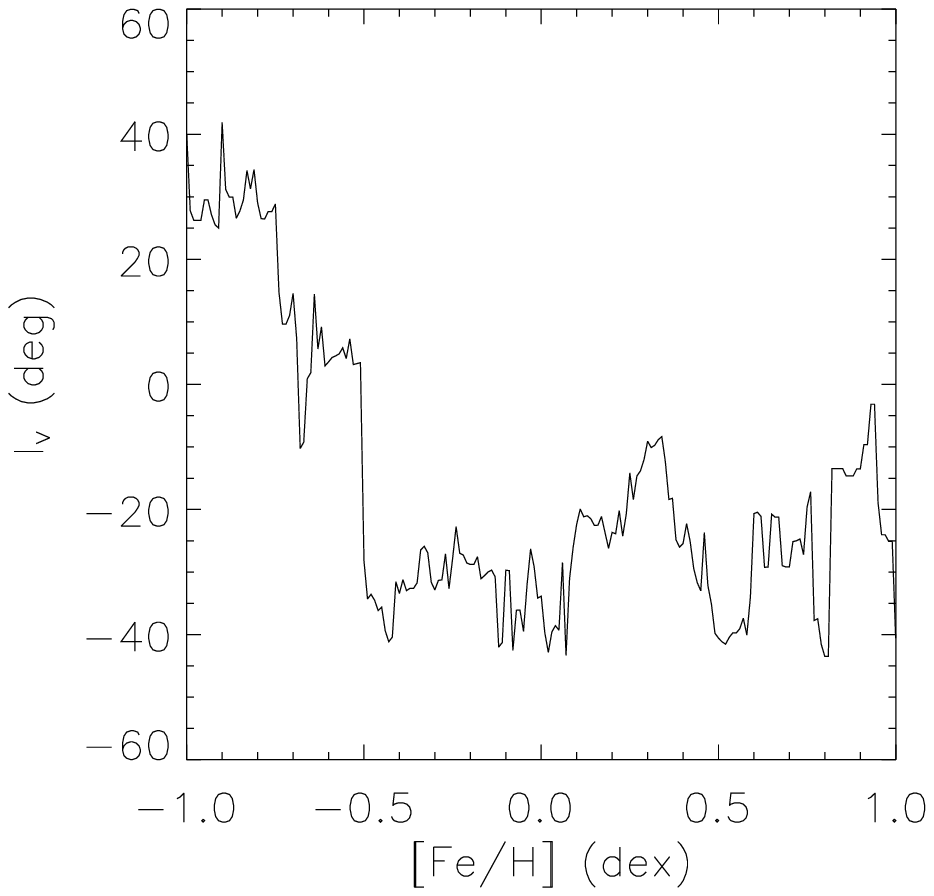}
\includegraphics[width=0.32\hsize]{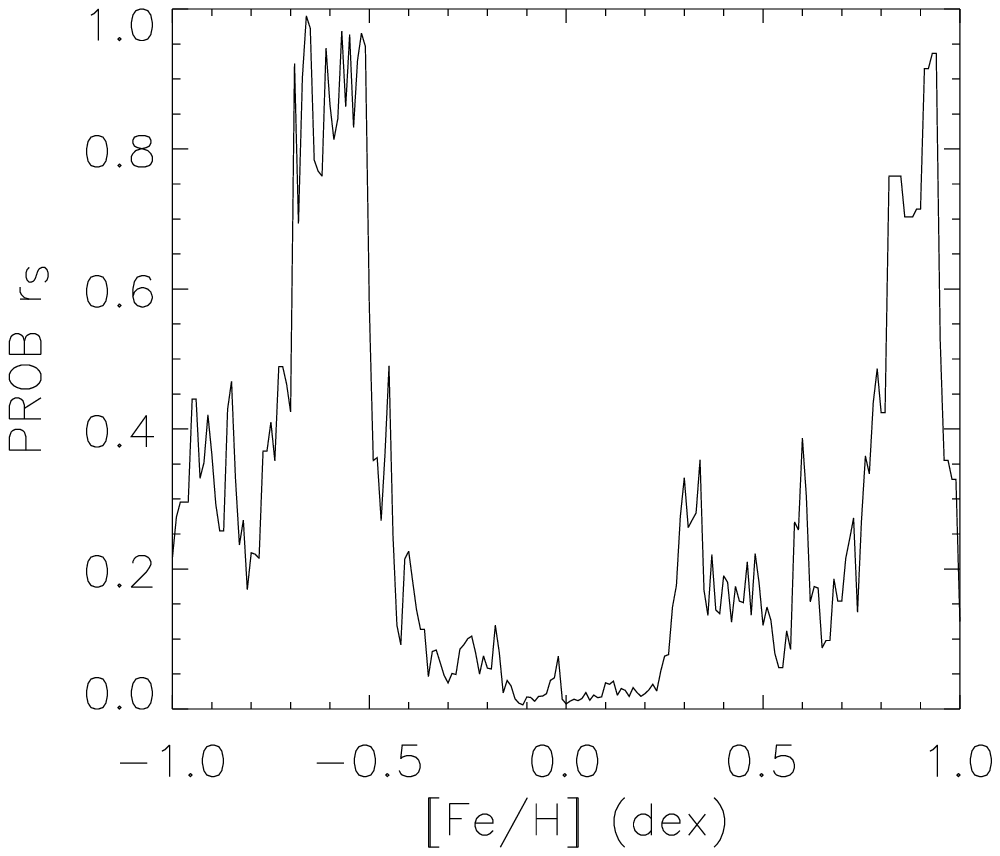}
\includegraphics[width=0.32\hsize]{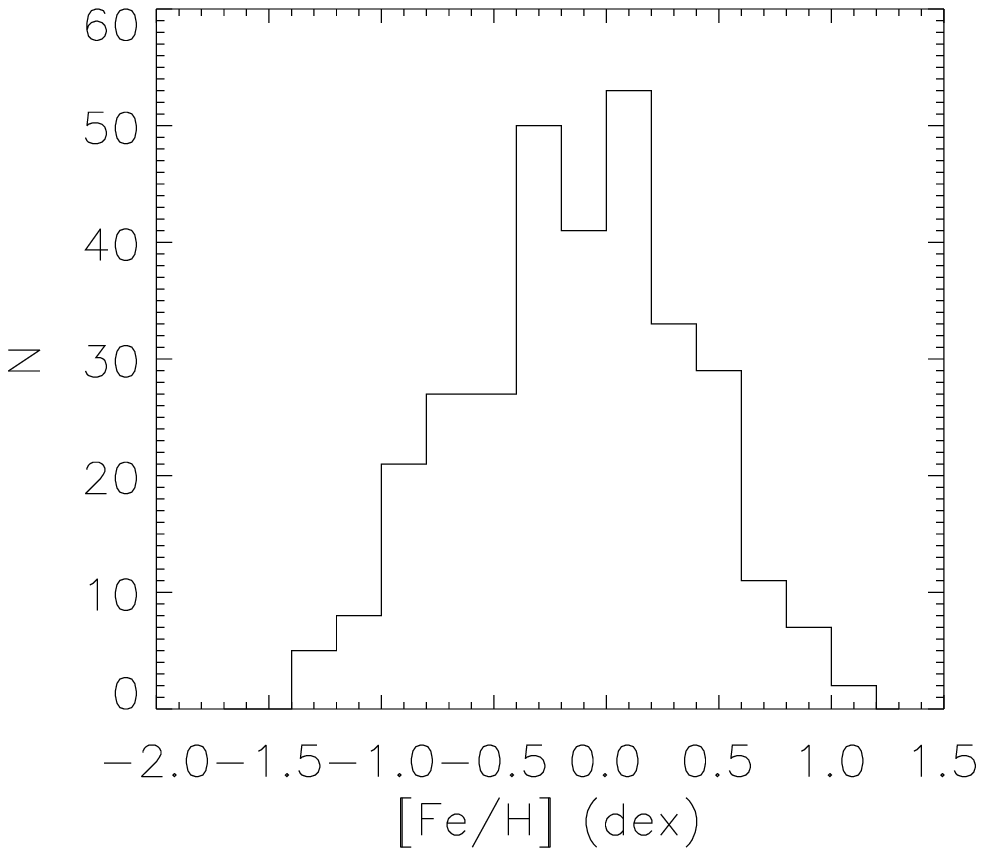}
\includegraphics[width=\hsize]{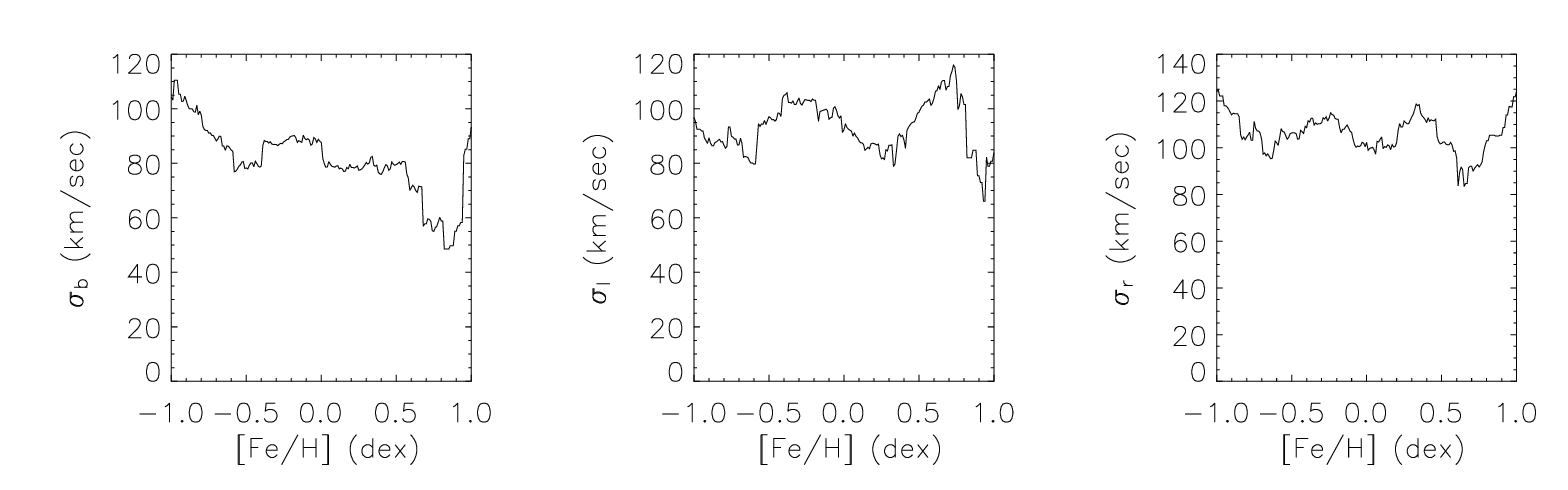}
\caption{The kinematics as a function of metallicity. \emph{Top row: Left},
  Vertex angle $l_v$ vs. $\rm [Fe/H]$ using a running box car of width 0.4
  dex. 
  \emph{Center}, the corresponding significance of the vertex
  deviation, using a non-parametric Spearman rank correlation. The
  combined significance for all stars with $\rm [Fe/H]>-0.5$ is greater
  than 99\%.
  \emph{Right}, histogram showing the number of stars in 0.2 dex
  $\rm [Fe/H]$ bins. \emph{Bottom row}, velocity dispersions in $b$, $l$ and
  $r$, for the same running metallicity bins. Note the break in the
  kinematics in $\sigma_b$ at $\rm [Fe/H]=-0.5$.
\label{fig:abunkin}}
\end{figure*}

The most significant vertex deviation is seen in
Fig.~\ref{fig:veldisp} for the metal rich sample ($\rm [Fe/H]>-0.5$),
whereas metal poor stars show no significant deviation. This result is
consistent with the results from \citet{zhao94}, based on
$\sim20\%$ of the present sample, and proves the reliability of their results
(Zhao et al.\ also saw a vertex deviation in the metal-poor sample, of
opposite sign to the metal-rich set: with the addition of the new
stars in this study the significance of this deviation has dropped).
This difference between metal rich and metal poor stars is a strong
stellar-dynamical demonstration of the existence of a stellar bar in
the galactic center \citep{binney,zhao94}, and suggests that much of
the disk takes part in it.

Fig.~\ref{fig:abunkin} shows the velocity dispersion components and
vertex deviation as a function of the metallicity, using a running
mean over metallicity bins of width 0.4dex. The Figure clearly shows a
rather sharp transition near $\rm [Fe/H]=-0.5$: more metal-poor stars
show little or no deviation, whereas the more metal-rich stars'
velocity ellipsoid is misaligned by about $30^\circ$, with a rank
correlation between radial and longitudinal velocity components for
these stars that is significant at the 99\% level. A further
indication of this change in kinematic properties is the dispersion in
the (completely independent) vertical velocity component, $\sigma_b$,
which shows a clear break at $\rm [Fe/H]=-0.5$ with the more
metal-poor stars' $b$ velocity dispersion being higher.

We have also examined the correlations between the $b$ velocity and
the other components, but found no significant deviation of the
velocity ellipsoids out of the plane.

The giant stars thus consist of a metal-rich population with a
significant vertex deviation of $30$ degrees, and a metal-poor
population that has no significant deviation, and hotter kinematics.

The general trend of the bulge to have this vertex deviation is
also seen in three minor axis fields with proper motions from
HST imaging with WFPC2 (Soto et al. 2007 in prep.).   
These fields were observed
for radial velocities 
using the integral field spectrograph on VIMOS at VLT.
The combination of radial velocities and $l$ proper motions
yields an observable vertex deviation in the fields (in samples
not segregated by metallicity).

\section{Discussion}

Our observations support the presence of a striking transition in the
kinematics of the bulge, from an apparently isotropic oblate spheroid
to a bar, at $\rm [Fe/H]=-0.5$.  We have used a non-parametric rank
correlation statistic to determine that the significance of the vertex
deviation for the metal-rich stars is $>99$\%.  There is a
corresponding but shallower trend toward lower vertical velocity
dispersion $(\sigma_b)$ at higher abundance.  It is difficult to
imagine how a system with such a kinematic transition could be
considered to have evolved in a one-zone chemical evolution scenario.

Do abundance trends of any other elements show a transition at $\rm
[Fe/H]=-0.5$?  Examining the trends given in \citet{fmr06b} and
\citet{lecu06}, we find no break in the trends of Mg, O, Ca, Si, Ti,
Na, and [Al/Fe] with [Fe/H] that would correspond to $-0.5$ dex.
\citet{fmr06b} consider the trend of the explosive alpha elements $\rm
\langle Ca+Si+Ti\rangle=\alpha_{exp}$ with [Fe/H].  They find that
even to the lower limit of bulge [Fe/H], at $\rm [Fe/H]<-0.5$, the
bulge trend of $[\alpha_{exp}/\rm Fe]$ lies above the locus of halo
and thick disk stars.  Hence, the stars with $\rm [Fe/H]<-0.5$ cannot
be considered simply to belong to the halo or thick disk.  Similarly,
virtually all stars with $\rm [Fe/H]>-0.5$ have strongly enhanced Mg,
a strong indication that the chemistry reflects a history of rapid
enrichment that extends through the boundary of this kinematic
transition.

In order to form the bar via a stellar-dynamical instability, which
then thickens into a bulge \citep{raha91}, the stars would already
have to have been in place.  The kinematics thus suggest a scenario in
which all stars with $\rm [Fe/H]>-0.5$ stars were formed in a disk
{\em before} the bar instability occurred; the more metal-poor stars
are then a population unrelated to the disk, bar and bulge, consistent
with their hotter kinematics.
But the continuity in bulge abundance trends across $\rm [Fe/H]=-0.5$
argues against the existence of two entirely separate populations with
different chemical evolution histories. Evidently the situation is
more complicated than a simple few-component model can describe. It is
clearly important to explore the existence of these trends in larger
samples and in additional fields in the bulge.

\acknowledgements

We thank the Lorentz Center in Leiden for an enjoyable workshop during
which this paper was completed.  RMR acknowledges support from GO-9816
from the Space Telescope Science Institute. KK acknowledges travel
support from the Leids Kerkhoven-Bosscha Fonds.

\end{document}